# Electronic properties and Fermi surface for new Fe-free layered superconductor BaTi$_2$Bi$_2$O from first principles

*D.V. Suetin,\* A.L. Ivanovskii*

*Institute of Solid State Chemistry, Ural Branch of the Russian Academy of Sciences, Ekaterinburg, GSP-145, 620990, Russia*

Very recently, as an important step in the development of layered Fe-free pnictide-oxide superconductors, the new phase BaTi$_2$Bi$_2$O was discovered which has the highest $T_C$ (~ 4.6 K) among all related non-doped systems. In this Letter, we report for the first time the electronic bands, Fermi surface topology, total and partial densities of electronic states for BaTi$_2$Bi$_2$O obtained by means of the first-principles FLAPW-GGA calculations. The inter-atomic bonding picture is described as a high-anisotropic mixture of metallic, covalent, and ionic contributions. Besides, the structural and electronic factors, which can be responsible for the increased transition temperature for BaTi$_2$Bi$_2$O (as compared with related pnictide-oxides BaTi$_2$As$_2$O and BaTi$_2$Sb$_2$O), are discussed.

**I. Introduction.** The discovery of superconductivity in layered Fe-*Pn* [1] and Fe-*Ch* [2] materials (where *Pn* and *Ch* are pnictogens and chalcogens, respectively), for which superconductivity emerges in [Fe$_2$*Pn*$_2$] or [Fe$_2$*Ch*$_2$] blocks, has stimulated intense search for new layered superconductors (SCs) and comprehensive investigations of their properties. Besides superconductivity, these materials exhibit a huge variety of exciting physical properties such as charge or spin density waves (CDW or SDW), different types of magnetic ordering, coexistence of SC and CDW/SDW phases, as well as coexistence of magnetism and superconductivity *etc*., see reviews [3-11].

An alternative line in the development of the new layered SCs is the search for Fe-free materials. Here, the layered pnictide-oxides BaTi$_2$Sb$_2$O (so-called 1221 phase) and Na$_2$Ti$_2$Sb$_2$O (so-called 2221 phase) have attracted considerable attention, when a low-$T_C$ superconducting transition at 1.2 K for BaTi$_2$Sb$_2$O [12] and emergence of superconductivity for hole-doped Ba$_{1-x}$Na$_x$Ti$_2$Sb$_2$O (up to $T_C$ ~ 5.5 K for x ~ 0.33 [13]) were found.

Very recently (2013, [14]), the newest 1221-like SC BaTi$_2$Bi$_2$O was discovered, for which the highest $T_C$ (~ 4.6 K) among all known non-doped 1221 and 2221 phases was established.

In order to get a detailed insight into the basic electronic properties and the peculiarities of the Fermi surface of this newest quaternary layered Fe-free pnictide-oxide SC, in this Letter we present first-principles calculations of BaTi$_2$Bi$_2$O. Our results cover the optimized lattice parameters and atomic positions, electronic bands, Fermi surface topology, as well as total and partial densities of electronic states. Additionally, Bader's analysis and the charge density isosurface are used in discussing the inter-atomic bonding for the examined material.

**II. Model and computational aspects.** The synthesized 1221 phase BaTi$_2$Bi$_2$O has [14] a tetragonal structure (space group *P*4/*mmm*) and can be described as blocks [Ti$_2$Bi$_2$O], which alternate with single sheets of Ba ions, Fig. 1. In turn, each block [Ti$_2$Bi$_2$O] consists of a Ti$_2$O square plane (with anti-CuO$_2$ configuration) placed between two Bi atomic layers, where Bi atoms are ranged symmetrically above and below the center of Ti squares. The crystal structure of BaTi$_2$Bi$_2$O is defined by the lattice parameters *a* and *c* and the internal parameter $z_{Bi}$.

Our calculations were performed by means of the full-potential linearized augmented plane wave method with mixed basis APW+lo (FLAPW) implemented in the WIEN2k suite of programs [15]. The generalized gradient approximation (GGA) to exchange-correlation potential in the well-known PBE form [16] was used. The basis set inside each muffin tin (MT) sphere was split into core and valence subsets. The core states were treated within the spherical part of the potential only and were assumed to have a spherically symmetric charge density in MT spheres. The valence part was treated with the potential expanded into spherical harmonics to *l* = 4. The valence wave functions inside the spheres were expanded to *l* = 12. We used the corresponding atomic radii: 2.10 a.u. for Ba, 1.90 a.u. for Ti, 2.10 a.u. for Bi, and 1.70 a.u. for O atoms. The plane-wave expansion was taken to $R_{MT} \times K_{MAX}$ equal to 7, and the *k* sampling with 12×12×6 *k*-points in the Brillouin zone was used. The self-consistent calculations were considered to be converged when the difference in the total energy of the crystal did not exceed 0.1 mRy as calculated at consecutive steps. The hybridization effects were analyzed using the densities of states (DOSs), which were obtained by the modified tetrahedron method [17], and some peculiarities of the chemical bonding picture were visualized by means of the charge density isosurface. To describe ionic bonding, a Bader's analysis [18] was carried out.

**III. Results and discussion.** At the first step, the equilibrium lattice constants for BaTi$_2$Bi$_2$O phases were calculated with full structural optimization including the internal parameter $z_{Bi}$. The results obtained are listed in the Table and are in reasonable agreement with the available experiment [14].

The near-Fermi electronic bands and total and partial densities of states (DOSs) of BaTi$_2$Bi$_2$O are depicted in Fig. 2. We see that the valence spectrum includes several characteristic bands: three lowest quasi-core bands separated by gaps and centered about -19.5 eV, -14.0 eV, and -10.9 eV below the Fermi level (E$_F$), which originate from O 2*s*, Ba 5*p*, and Bi 6*s* states, respectively, and a broad near-Fermi band with the bandwidth of about 7.23 eV. In turn, this band includes a subband of hybridized (O 2*p* - Ti *s,p,d*) states (centered at -5.9 eV), which are responsible for covalent O-Ti bonds, a subband of hybridized (Bi 6*p* - Ti 3*d*) states (centered at -2.2 eV), which form directional Bi-Ti bonds, and a subband composed mainly of Ti 3*d* states, which directly adjoin E$_F$. Note that the contributions to the near-Fermi region from the sheets of Ba ions are very small (see also the Table). Thus, conduction in this material will be anisotropic and will happen mainly in blocks [Ti$_2$Bi$_2$O].

The calculated total density of states at the Fermi level, $N(E_F)$ = 3.418 states/eV·form.unit, allows us to estimate (within the free electron model) the Sommerfeld constant $\gamma$ = 8.06 mJ·K$^{-2}$·mol$^{-1}$ and the Pauli paramagnetic susceptibility $\chi$ = 1.10 ·10$^{-4}$ emu/mole. Note that these values are comparable with the same for the layered Fe-containing SCs such as multi-component pnictide-oxides $Sr_2ScFeAsO_3$ ($\gamma$ = 8.85 mJ·K$^{-2}$·mol$^{-1}$ [19]) or $Sr_3Sc_2Fe_2As_2O_5$ ($\gamma$ = 6.77 mJ·K$^{-2}$·mol$^{-1}$ [20])

Since electrons near the Fermi level are involved in the formation of the superconducting state, it is important to understand their nature. The atomic decomposed partial DOSs at the Fermi level are listed in the Table. It is seen that the main contribution to $N(E_F)$ comes from the Ti $3d$ states, which are responsible for the metallic-like properties of this material. Among them, the $d_{xz}$ and $d_{yz}$ orbitals give very little contributions to $N(E_F)$. These orbitals are directed towards Bi atoms forming (Ti $3d$ – Bi $6p$) covalent bonds and are shifted above and below of the Fermi level. On the contrary, the three other orbitals: $d_{xy}$ (which form metal–metal bonds between the neighboring Ti atoms), $d_{x2-y2}$ (directed to $O^{2-}$ ions), and $d_{z2}$ (directed along the $c$ axis) give the main contributions to $N(E_F)$.

These three orbitals participating in the formation of the quasi-flat near-Fermi bands define the main features of the Fermi surface (FS), Fig. 1. The FS of $BaTi_2Bi_2O$ adopts a multi-sheet type and contains four disconnected hole-type and electron-type pockets. Among them, two 2D-like hole- and electron-type sheets in the form of deformed cylinders (on the lateral sides and in the corners of the BZ, respectively) are extended along the $k_z$ direction. Besides, two closed hole- and electron-like pockets of complicated rosette-like forms are centered at the $Z$ and $\Gamma$ points, respectively, Fig. 1.

According to our calculations, the common picture of inter-atomic interactions for $BaTi_2Bi_2O$ can be described as a high-anisotropic mixture of metallic, covalent, and ionic contributions, where the *metallic-like* bonding is due mainly to the interaction of Ti $3d_{xy}$ orbitals of neighboring Ti atoms inside blocks [Ti$_2$Bi$_2$O]. Numerically, this metallic-like bonding can be estimated using the so-called metallicity parameter [21] $f_m$ = 0.026$N(E_F)/n_e$, where $n_e$ = N/V; N is the total number of valence electrons; and $V$ is the cell volume. The calculated value for $BaTi_2Bi_2O$ is $f_m$ = 0.496. Again, this value is comparable, for example, with $f_m$ for the aforementioned Fe-containing layered oxide $Sr_4Sc_2Fe_2As_2O_6$, $f_m$ = 0.36 [20,22], but is smaller than for the other 2221-like Fe-containing oxide $Na_2Fe_2Se_2O$, $f_m$ = 1.036 [23].

In turn, *covalent* bonds arise also inside blocks [Ti$_2$Bi$_2$O] as a result of hybridization of Ti-Bi and Ti-O valence states (see above), and this in-block covalent bonding, as well as *ionic* bonding between blocks [Ti$_2$Bi$_2$O] and atomic Ba sheets are well visible in Fig. 1, where the electron density iso-surface is depicted. As to ionic bonding, the widely used simplified ionic model assuming the oxidation numbers of atoms gives immediately the ionic formula $Ba^{2+}Ti^{3+}{}_2Bi^{3-}{}_2O^{2-}$ or, in the other form, $Ba^{2+}[Ti_2Bi_2O]^{2-}$. The actual atomic charges (as obtained within the Bader's model [18]) differ from the formal ionic charges $Ba^{1.18+}Ti^{1.27+}{}_2Bi^{1.18-}{}_2O^{1.35-}$ owing to covalence in blocks [Ti$_2$Bi$_2$O] and the charge transfer between the adjacent atomic Ba sheets and the blocks [Ti$_2$Bi$_2$O]: $\delta Q$ = 1.18 *e.*

To understand the peculiarities of the newly discovered superconducting phase BaTi$_2$Bi$_2$O and note some factors, which can be responsible for its increased transition temperature, let us discuss the trends in structural and electronic properties in a series of isostructural phases BaTi$_2$As$_2$O → BaTi$_2$Sb$_2$O → BaTi$_2$Bi$_2$O, Fig. 3.

For these phases, charge- or spin-density wave (CDW/SDW) states appear, with CDW/SDW transitions at $T_a$ ~ 200 K for the non-superconducting BaTi$_2$As$_2$O and at $T_a$ ~ 50 K for BaTi$_2$Sb$_2$O (with $T_C$ ~ 1.2 K) [24], whereas for the newly discovered phase BaTi$_2$Bi$_2$O (with the highest $T_C$ ~ 4.6 K) the CDW/SDW state disappears [14]. This effect is related to decreased overlap between orbitals of the nearest Ti-Ti atoms, which should destabilize the CDW/SDW state, see [12-14,24,25]. Indeed, in the sequence BaTi$_2$As$_2$O →...→ BaTi$_2$Bi$_2$O, the parameter *a* (*i.e.* the Ti-Ti distance $d^{Ti-Ti} = a/\sqrt{2}$) grows (Fig. 3), and for BaTi$_2$Bi$_2$O the calculated value $d^{Ti-Ti}$ = 2.9149 Å (2.916 Å as obtained in experiment [14]) exhibits the maximal deviation from $d^{Ti-Ti}$ = 2.87 Å for Ti metal leading to the maximal reduction of Ti-Ti orbital overlap, which is favorable for superconductivity.

The second (electronic) factor will be discussed by comparing the near-Fermi electronic states of BaTi$_2$Bi$_2$O with those for BaTi$_2$As$_2$O and BaTi$_2$Sb$_2$O (Fig. 3), as calculated by the authors within the same FLAPW-GGA approach. Since the electron-phonon coupling mechanism was found applicable for superconductivity in the related phase BaTi$_2$Sb$_2$O [26], the simple correlation $T_C$ ~ $N(E_F)$ should be expected. However, from Fig. 3 we see that $N(E_F)^{BaTi2Bi2O}$ ~ $N(E_F)^{BaTi2As2O}$ < $N(E_F)^{BaTi2Sb2O}$. Thus, we can speculate that the growth of the transition temperature in the sequence BaTi$_2$As$_2$O →...→ BaTi$_2$Bi$_2$O should be due mainly to the structural factor, when isovalent atomic substitutions of pnictogens with larger radii ($R(As^{3-}) < R(Sb^{3-}) < R(Bi^{3-})$) result in progressing suppression of CDW/SDW instability.

**IV. Conclusions.** In summary, by means of the first-principles calculations, we studied in details the electronic properties, Fermi surface topology, and the inter-atomic bonding picture for the very recently discovered layered Fe-free pnictide-oxide superconductor BaTi$_2$Bi$_2$O with the highest $T_C$ (~ 4.6 K) among all related non-doped systems. We found that the near-Fermi electronic bands, which are involved in the formation of superconducting state, arise mainly from the Ti 3*d* states of the blocks [Ti$_2$Bi$_2$O]. In turn, among these Ti 3*d* states, the orbitals $d_{xy}$, $d_{x2-y2}$, and $d_{z2}$ give the main contributions to this region and define the features of the Fermi surface, which adopts a multi-sheet type and combines four disconnected hole- and electron-like pockets. The inter-atomic bonding in BaTi$_2$Bi$_2$O is high-anisotropic and includes a mixture of covalent, ionic, and metallic contributions inside blocks [Ti$_2$Bi$_2$O], whereas the bonding between blocks [Ti$_2$Bi$_2$O] and atomic Bi sheets is of an ionic type.

Finally, our analysis reveals that the growth of $T_C$ for BaTi$_2$Bi$_2$O (in comparison with related pnictide-oxides) should be due mainly to the structural factor, *i.e.* progressing suppression of CDW/SDW instability. On the other hand, the electronic factor can also promote a further growth of $T_C$ - for example, for the proposed Ba$_{1-x}$Na$_x$Ti$_2$Bi$_2$O [14]. In this case, as can be seen in Fig. 2, the Fermi level for the hole-doped Ba$_{1-x}$Na$_x$Ti$_2$Bi$_2$O should be shifted downward to the peak with increased N(E$_F$), which is favorable for the growth of $T_C$.

* suetin@ihim.uran.ru

**Calculated lattice constants ($a$ and $c$, in Å), atomic positions, total and partial DOSs at the Fermi level (in states/eV/form.unit) for BaTi$_2$Bi$_2$O as obtained within FLAPW-GGA**

| atomic positions | lattice constants * | DOSs at the Fermi level | |
|---|---|---|---|
| Ba(1$d$): ½; ½0; ½;<br>Ti(2$f$): ½; 0; 0;<br>Bi(2$g$): 0; 0; $z_{Bi}$ **<br>O(1$c$): ½; ½; 0 | $a$ = 4.1220 (4.1232)<br>$c$ = 8.5467 (8.3447) | total: 3.418<br>Ti 3$d$: 1.646<br>Bi 6$p$: 0.127<br>(O + Ba) = 0.069 | Ti $d_{xy}$: 0.573<br>Ti $d_{x2-y2}$: 0.357<br>Ti $d_{z2}$: 0.687 |

* available experimental results [14] are given in parentheses.
** $z_{Bi}$ = 0.7477 (our calculations) and $z_{Bi}$ = 0.7487 (experiment [14]).

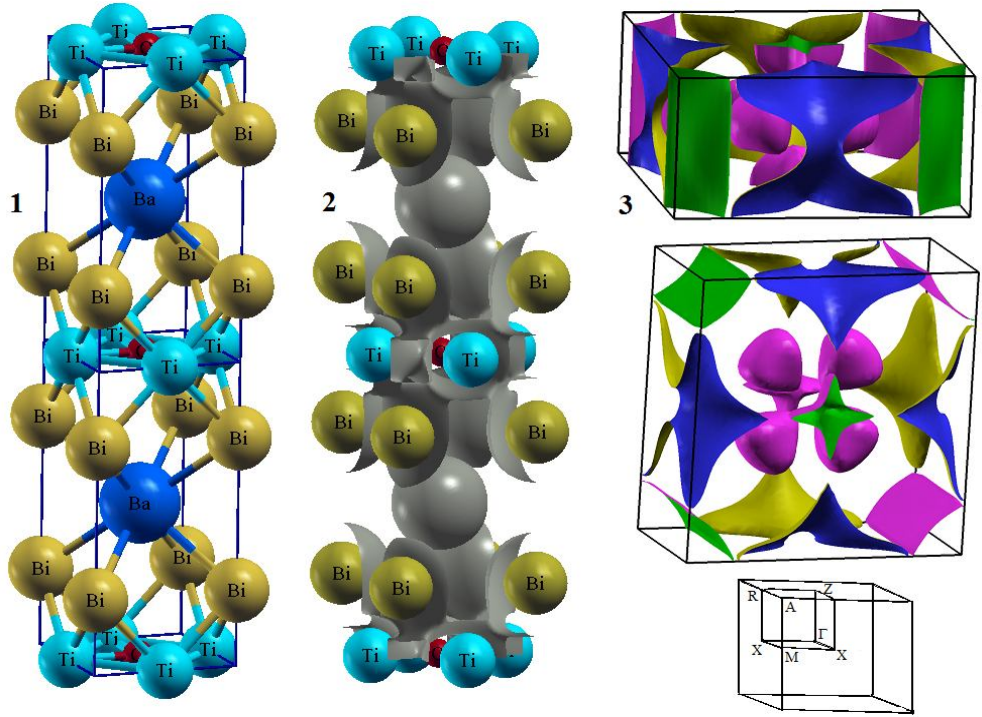

**Fig. 1.** Crystal structure of BaTi$_2$Bi$_2$O (1); Isosurface of electronic charge density ($\rho$=0.2 e/Å$^3$) (2), and Fermi surface (*side* and *top* views), at the bottom is the Brillouin zone (3).

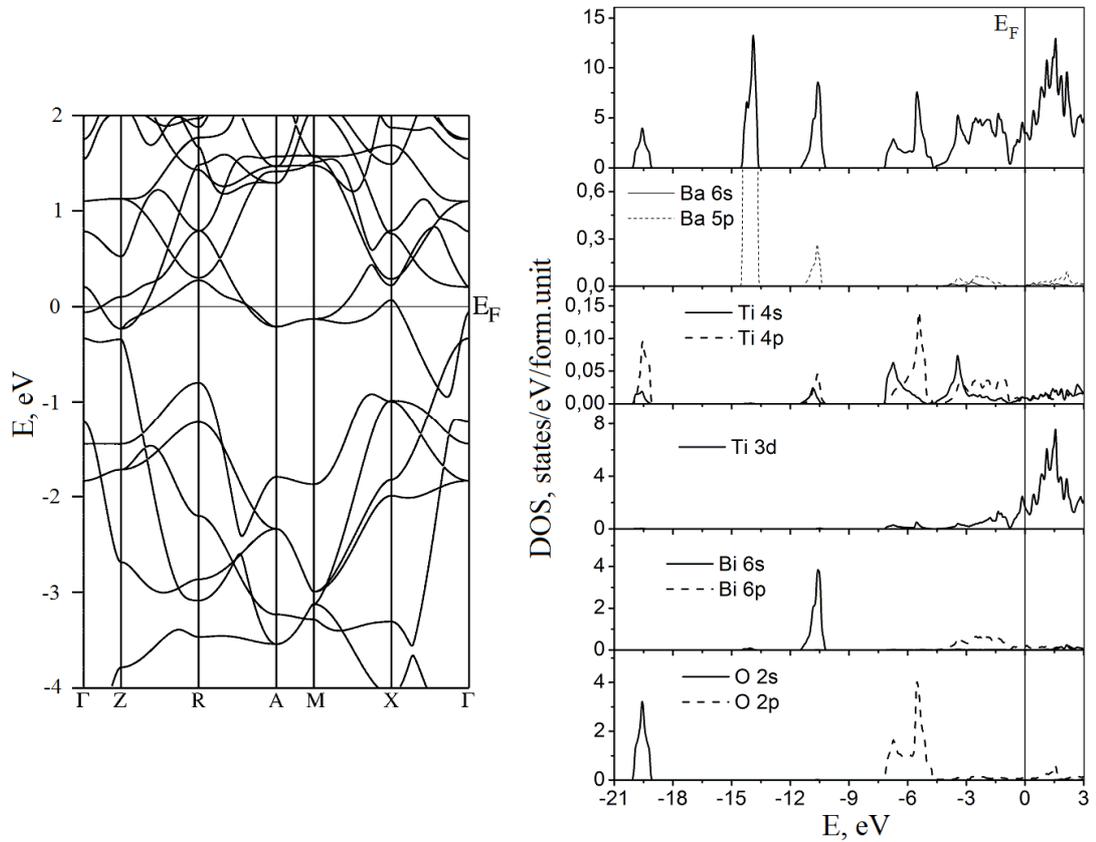

**Fig. 2.** Calculated near-Fermi electronic bands (1) and total and partial densities of states for BaTi$_2$Bi$_2$O (2).

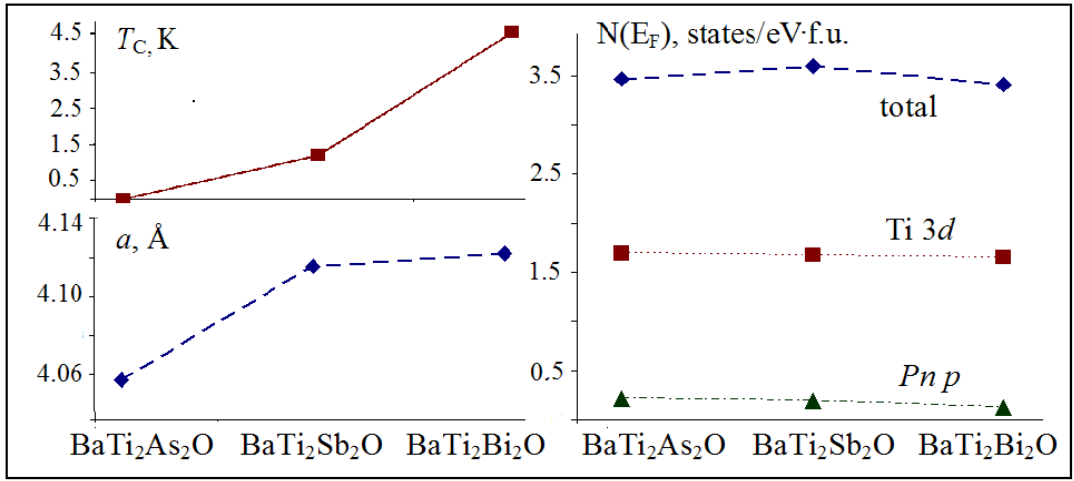

**Fig. 3.** Changes in transition temperature $T_C$, lattice parameter $a$, and total Ti 3$d$- and $Pn$ $p$ states ($Pn$ = As, Sb and Bi) at the Fermi level in the sequence BaTi$_2$As$_2$O → BaTi$_2$Sb$_2$O → BaTi$_2$Bi$_2$O as obtained within FLAPW-GGA.